# Self-Consistent Local Mean-Field Theory for Phase Transitions and Magnetic Properties of FeRh

Brianne McGrath, Robert E. Camley, Karen L. Livesey

Center for Magnetism and Magnetic Nanostructures

University of Colorado at Colorado Springs

1420 Austin Bluffs Parkway

Colorado Springs, CO 80918

**Abstract**

FeRh has a phase transition from an antiferromagnetic state (low temperature) to a ferromagnetic state (high temperature) at 360 K. Various explanations for this behavior have been proposed over the past 20 years. However, many of the mechanisms are inconsistent with all the data. Early models were Ising-like, but the large anisotropy fields necessary for this are not found in hysteresis curves. Using a four-spin Hamiltonian, we obtain a complete theoretical description of the field and temperature phase diagram and the magnetic properties for FeRh. The theoretical results are in good agreement with experiments. We use our approach to predict changes in behavior as a function of the thickness of an FeRh film. We find the four-spin Hamiltonian is particularly sensitive to the presence of a surface, with thinner films remaining ferromagnetic over a larger temperature range because the four-spin contribution to the energy (which favors the antiferromagnetic state) is smaller.

## I. INTRODUCTION

Discovered in 1938[1,2], FeRh has recently attracted much attention due to its interesting magnetic properties and potential applications. FeRh (CsCl structure) has a paramagnetic-ferromagnetic phase transition around 650 K.[2,3] More motivating is the interesting antiferromagnetic (AFM) to ferromagnetic (FM) phase transition upon heating just above room temperature, near 360 K. This transition is shown for a thin film in Fig. 1, where a schematic shows the magnetic moments (arrows) associated with Fe and Rh ions on the lattice.

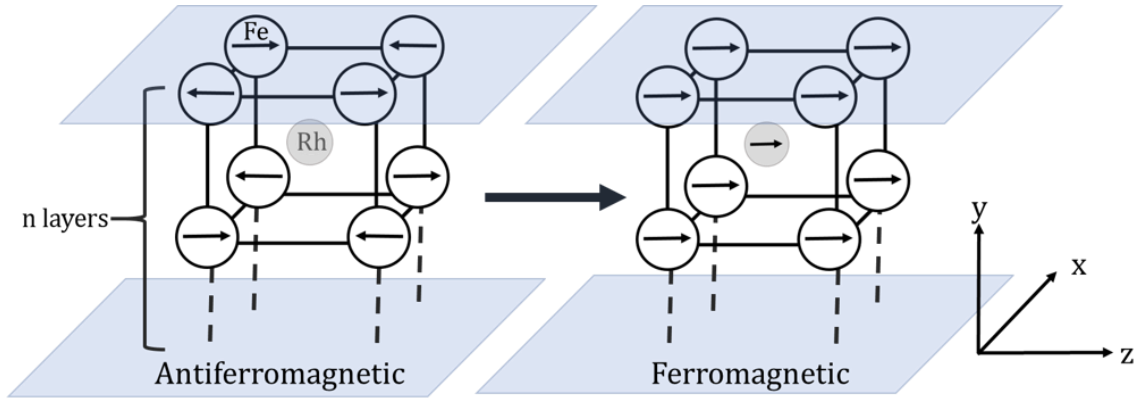

FIG 1. Illustration of the phase transition of a thin film of FeRh with a surface on the top and bottom of the film. The CsCl structure of the unit cell is shown. The moments of the Fe are antiferromagnetically arranged at low temperatures (left) and transition to a ferromagnetic state at higher temperatures (right). The Rh atom gains a small moment in the ferromagnetic state. The moments are assumed to lie in plane due to the demagnetization field.

Around and at this phase transition other interesting physical properties can be seen including a strong magnetocaloric effect,[4–6] a 0.9 percent volume expansion,[7–10] a drop in resistivity leading to a huge magnetoresistance,[8,11] and an ultrafast phase transition.[12–14] In addition, the transition temperature range can easily be fine-tuned by changing the composition slightly,[15] altering preparation[3] and annealing conditions[16] or changing the strain,[17] doping,[15,18,19] magnetic field,[20] stress,[21] terminations,[22,23] substrate interfaces,[24–26] and thickness.[17,22] Mainly due to the AFM to FM phase transition near room temperature, FeRh has potential applications in magnetic memory and recording media.[12,15,24,27]

Although the magnetic properties of bulk FeRh have been extensively studied, the mechanism behind the AFM to FM transition is still widely debated and not well understood. It has been suggested that the transition could be structural, due to the well-known ~1% volume expansion of the unit cell in the FM phase,[7-10] or it could be purely magnetic.[13] It has additionally been suggested that the phase transition can be laser induced,[9,12] or have a contribution from spin waves.[28] Moreover, the specific role of Rh is not well understood, but is believed to have a significant influence on the phase transition.[18,29,30]

Some of the early, purely-magnetic models relied on large values of magnetic anisotropy in order to reproduce the magnetic behavior. In particular, Ising-like models[24,31] could obtain the FM to AFM phase transition at an appropriate temperature. However, these calculations would also imply significant values for the coercive field, something not seen in experiments. More recently, atomistic spin dynamic methods have been used to study the phase transition.[13,17] However, Barker and Chantrell's work[13] found the transition temperature in bulk only. Similarly, the results of Ostler *et al.*[17] do not include information regarding magnetic field dependence. Our work will therefore address some of the situations not covered in these works. We also use a method that is computationally less-expensive than atomistic methods, and includes realistic anisotropy energy contributions.

In this paper, we use a self-consistent local mean field theory[24,32] with a higher-order four-spin contribution to the exchange energy. We are able to obtain a complete, purely magnetic, theoretical description of the temperature and field phase diagrams for this system, which are in good agreement with experiments and some other theoretical models. In addition, we study the effect of various parameters, such as thickness, applied field and interface/surface effects, on the transition temperature and coercive field of FeRh.

We obtain results for the following properties of FeRh:

1. Bulk magnetization versus temperature (M vs T) and magnetization versus applied field (M vs H) behavior, including thermal hysteresis;
2. Effect of applied field on the AFM to FM transition temperature range for bulk FeRh;
3. Thickness dependence of both M vs H and M vs T behavior of a thin film of FeRh;
4. Penetration depth for surface induced changes in the magnetic structure as a function of temperature and thickness.

Our model produces a comprehensive description of FeRh ranging from bulk to ultrathin films, and we discuss in detail our results as compared to experimental findings. We see similar behavior to that of experiments in that the AFM to FM transition temperature is reduced as the film thickness is reduced.[33-35] Moreover, a thermal hysteresis in bulk-like films of 30-40 K is found, in agreement with experiments.[33] Furthermore, we reproduce the lowering of the AFM to FM transition temperature with increasing applied field and importantly are able to visualize the structure of the moments during these transitions. Through this we are able to understand the physical behavior, such as canting of the spins, of the system under different conditions.

## II. Theoretical Considerations

We use the following energy for an Fe atom associated with site *i*, in FeRh:

$$\mathcal{E}(i) = \mathcal{E}_{Zeeman}(i) + \mathcal{E}_{ex}(i) + \mathcal{E}_{anis}(i). \quad \textbf{(1)}$$

We note that the Rh is not explicitly included in our model. This is not an issue in the AFM state where Rh has zero moment. Once the system transitions to the FM state, Rh gains a moment of $0.9\mu_B$ due to polarization from the surrounding Fe atoms.[29] The major effect of neglecting the Rh atom is that the contribution from Rh to the total magnetization in the FM state is neglected. This is discussed further in the results section.

The first term in Eq. (1) is the Zeeman energy, the energy of the spin in the presence of an applied magnetic field **H**, and is given by

$$\mathcal{E}_{zeeman}(i) = -\mu_{Fe} \boldsymbol{s_i} \cdot \boldsymbol{H}, \quad (2)$$

where $\mu_{Fe} = 3.3\mu_B$ is the net magnetic moment of an iron site, which has been determined from neutron diffraction experiments[10,29] as well as self-consistent total-energy calculations,[36,37] and $\boldsymbol{s_i}$ is a classical spin vectors of length 1 associated with site *i*.

The $\mathcal{E}_{ex}$ term in Eq. (1) is the exchange energy, which is composed of the bilinear and higher-order non-Heisenberg exchange terms, namely $\mathcal{E}_{ex} = \mathcal{E}_{bilinear} + \mathcal{E}_{4\ spin}$. This expansion of the typical Heisenberg Hamiltonian was originally purposed by Mryasov[38] for metamagnetic phase transitions and first applied to FeRh by Barker and Chantrell[13]. The full expansion of the exchange energy is given by:

$$\mathcal{E}_{ex}(i) = -\sum_{nn} J_1 \boldsymbol{s_i} \cdot \boldsymbol{s_j} - \sum_{nnn} J_2 \boldsymbol{s_i} \cdot \boldsymbol{s_j} + \frac{1}{3} \sum_{quartets} D_{ijkl}[(\boldsymbol{s_i} \cdot \boldsymbol{s_j})(\boldsymbol{s_k} \cdot \boldsymbol{s_l})$$

$$+ (\boldsymbol{s_i} \cdot \boldsymbol{s_k})(\boldsymbol{s_j} \cdot \boldsymbol{s_l}) + (\boldsymbol{s_i} \cdot \boldsymbol{s_l})(\boldsymbol{s_k} \cdot \boldsymbol{s_l})]. \quad (3)$$

The first two terms are the bilinear exchange terms between Fe-Fe nearest neighbor (*nn*) and Fe-Fe next-nearest neighbor (*nnn*) pairs, where $J_1$ and $J_2$ are the respective exchange constants.

The third term in Eq. (3) is the four-spin, non-Heisenberg term, as given by Barker and Chantrell,[13] with $D_{ijkl}$ the four spin exchange constant. The sum is over all 32 basic quartets of the simple cubic lattice that include site *i*, four of which are shown in Fig. 2 with site *i* at the bottom left corner. As mentioned earlier, the energy involves the Fe spins only. The effect of the induced Rh moment and its coupling to the Fe spins are included in the higher order, Fe-Fe four-spin exchange coupling. The four-spin exchange term can be obtained in a perturbative expansion of the Hubbard model.[39] Normally these higher order interactions are magnitudes smaller than the Heisenberg term and therefore ignored, but they become more prevalent when 4d elements are present.[40,41] We use the same sign conventions for this term as Barker and Chantrell,[13] though it should be noted that often the four-spin Hamiltonian is written with differing signs.[39,42]

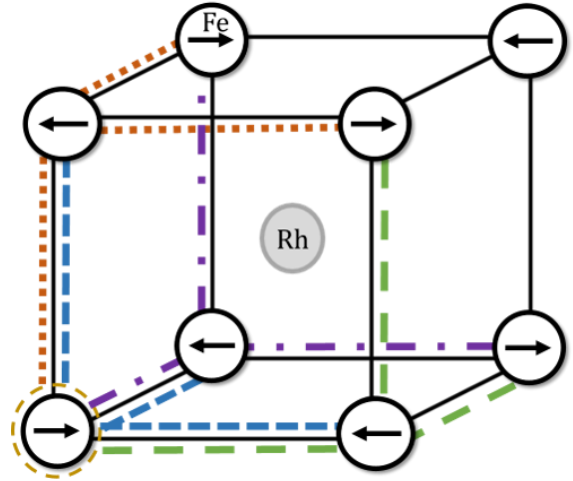

FIG 2. Illustration of the 4 basic quartets in the unit cell. Site I (lower left) is circled, and each color (pattern) represents one basic quartet of the 4 spin interaction in a unit cell. Each site has 32 basic quartets within a thin film.

The third term in Eq. (1) is a combination of the in-plane and out-of-plane anisotropies:

$$\varepsilon_{anis}(i) = -K_\parallel s_z^2(i)s_x^2(i) - K_\perp(i)s_y^2(i), \quad \textbf{(4)}$$

where and $K_\parallel$ and $K_\perp$ are the effective in-plane and out-of-plane anisotropy constants respectively, with *y* the out-of-plane direction for thin films. There is not much information on the in-plane anisotropy of FeRh. Some authors assume very large anisotropy (on the order of several $k_BT$),[24] while most state that FeRh is very soft and ignore it all together. All earlier theoretical models do not distinguish between in-plane and out-of-plane anisotropy. The first full experimental study done on the anisotropy of FeRh was by Mancini[43] in 2013 , which gives a large out-of-plane anisotropy and a much smaller in-plane anisotropy. A four-fold anisotropy is chosen here for the in-plane term with $K_\parallel = 1.18 \times 10^{-17}$ ergs in order to reasonably match experimental results by Mancini *et al*[43]. FeRh has a very large out of plane anisotropy which can be thought to effectively

6simple cubic lattice that include site *i*, four of which are shown in Fig. 2 with site *i* at the bottom left corner. As mentioned earlier, the energy involves the Fe spins only. The effect of the induced Rh moment and its coupling to the Fe spins are included in the higher order, Fe-Fe four-spin exchange coupling. The four-spin exchange term can be obtained in a perturbative expansion of the Hubbard model.[39] Normally these higher order interactions are magnitudes smaller than the Heisenberg term and therefore ignored, but they become more prevalent when 4d elements are present.[40,41] We use the same sign conventions for this term as Barker and Chantrell,[13] though it should be noted that often the four-spin Hamiltonian is written with differing signs.[39,42]

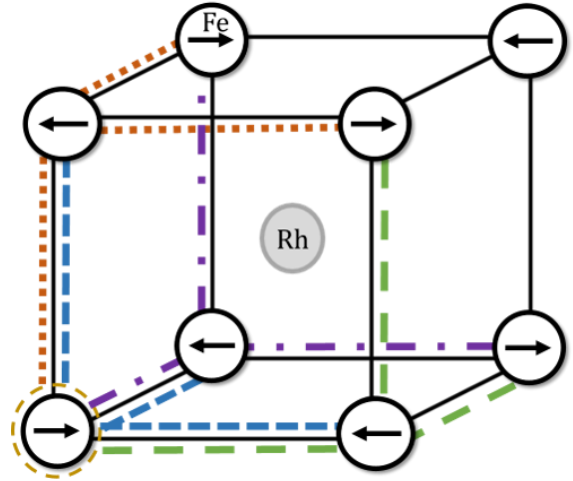

FIG 2. Illustration of the 4 basic quartets in the unit cell. Site I (lower left) is circled, and each color (pattern) represents one basic quartet of the 4 spin interaction in a unit cell. Each site has 32 basic quartets within a thin film.

The third term in Eq. (1) is a combination of the in-plane and out-of-plane anisotropies:

$$\varepsilon_{anis}(i) = -K_\parallel s_z^2(i)s_x^2(i) - K_\perp(i)s_y^2(i), \quad \textbf{(4)}$$

where and $K_\parallel$ and $K_\perp$ are the effective in-plane and out-of-plane anisotropy constants respectively, with *y* the out-of-plane direction for thin films. There is not much information on the in-plane anisotropy of FeRh. Some authors assume very large anisotropy (on the order of several $k_BT$),[24] while most state that FeRh is very soft and ignore it all together. All earlier theoretical models do not distinguish between in-plane and out-of-plane anisotropy. The first full experimental study done on the anisotropy of FeRh was by Mancini[43] in 2013 , which gives a large out-of-plane anisotropy and a much smaller in-plane anisotropy. A four-fold anisotropy is chosen here for the in-plane term with $K_\parallel = 1.18 \times 10^{-17}$ ergs in order to reasonably match experimental results by Mancini *et al*[43]. FeRh has a very large out of plane anisotropy which can be thought to effectively reduces the strength of the demagnetization field; therefore, we scale down the demagnetization field (in CGS units) from $4\pi M$ to $3\pi M$, as is done also in Mancini *et al.*[43] This will be discussed in detail later.

Our choice of values for $J_1$, $J_2$, and $D_{ijkl}$ are determined by the known Curie temperature and AFM to FM transition temperature of FeRh and can be found in Table I. It is well known that mean field theory provides both quantitatively and qualitatively correct results for phase transitions, but using effective exchange constants that are too small, especially in lower dimensions.[44] This is a typical feature of mean field theories because their neglect of fluctuations causes them to overestimate the tendency to order. In addition, our mean-field theory neglects the creation of possible domain walls laterally throughout the sample. In order to obtain quantitatively correct transitions, our parameters have been reduced (see Table I) from those used for atomistic theories which include fluctuations, being roughly half to one third of those values. It is important to note that the ratios of the exchange constants are, however, consistent, indicating that the competition between the various exchange interactions drive the AFM to FM transition in the same way in mean-field theories and in atomistic theories.

|       | Ours [ergs]         | Ratios              | Barker & Chantrell [ergs] | Ratios |
|-------|---------------------|---------------------|---------------------------|--------|
| $J_1$ | $1.7 \times 10^{-15}$  | $J_1/J_2 = 0.157$   | $4.0 \times 10^{-15}$        | 0.145  |
| $J_2$ | $10.8 \times 10^{-15}$ | $J_2/D = 13.01$     | $27.5 \times 10^{-15}$       | 11.95  |
| $D$   | $0.83 \times 10^{-15}$ | $J_1/D = 2.048$     | $2.3 \times 10^{-15}$        | 1.75   |

Table I: Comparison of the exchange constants used in this work with those of Ref. 13. Ratios of the constants are given for both this work and Ref. 13, and it should be noted they are consistent.

Implementing self-consistent local mean field theory, we created a multi-layered system, which can range from a few monolayers (ML) to a bulk material. Each layer has two unique Fe sublattices, allowing for an antiferromagnetic or canted state. Within mean field theory, it is considered that all of the spins on a single sublattice within a layer have the same thermal average and point in the same direction. (There is translational invariance in the film plane.) As a result, this calculation cannot give information regarding lateral domain formation, but does give insight to the thickness dependence. This model represents a pure, single crystal FeRh. It does not, for example, include the effect of a substrate, although this could be easily added. In addition, our exchange and anisotropy values are constant with temperature, similar to other theoretical treatments.

The magnetic system is described by Eq. (1). The effective magnetic field on each lattice site is found by:

$$\boldsymbol{H}_i = -\frac{1}{\mu_s}\frac{\partial \mathcal{E}}{\partial \boldsymbol{s}_i} \ . \qquad (5)$$

The thermal averaged spin is calculated using:

$$<\boldsymbol{s}_i> = s_i B_s(x) \quad , \qquad (6)$$

where $s_i$ is the spin on site i, which is 1 for all spins, and $B_S(x)$ is the Brillouin function defined by

$$B_s(x) = \frac{2s+1}{2s}\coth\left(\frac{2s+1}{2s}x\right) - \frac{1}{2s}\coth\left(\frac{x}{2s}\right) \qquad (7)$$

and the argument is given by

$$x = \frac{\mu_{Fe}\boldsymbol{S}_i \cdot \boldsymbol{H}_i(<\boldsymbol{s}_i>)}{k_B T}. \qquad (8)$$

Here $k_B$ is the Boltzmann constant. $\boldsymbol{H}_i(<\boldsymbol{s}_i>)$ is the effective field acting on site *i* at a given temperature. Special care must be taken for the dot product in the argument to avoid over-counting and to recover the correct energy.

We consider a spin in an arbitrary layer on a particular sublattice. The thermal averaged magnitude is found using Eq. (6) and the effective field is then calculated using Eq. (5). The spin is then rotated in the direction of the calculated effective field, lowering the energy. This process is done for both Fe sub-lattices on each layer and repeated for every layer in the system. The entire process is repeated with the newly calculated values until self-consistency of the entire system is reached, i.e. every spin is pointed along its local field and its magnitude and direction no longer change. At this point, the components of all the spins are recorded, and the magnetization of the system is determined. This method can allow the system to fall into a local energy minimum.[32] To check this, we started the system in multiple configurations at a given temperature and field. In general, away from the hysteretic regions we found only one final configuration.

To obtain the temperature dependence of the system, we use the final self-consistent state from a previous nearby temperature as the initial state at the new temperature. However, this can cause a problem in that the system can become trapped in a local energy minimum, resulting in nonphysical behavior where there is no phase transition or there is an extra-large hysteresis. To overcome this we modify the program by wiggling the spins in the initial state by a small amount, for them to get over the energy barrier and to avoid false, numerical stability. Typically the wiggle is less than 3 degrees in plane near the transition, but can be varied in size (0.5 to 6 degrees) without altering the results. To determine the magnetization of the system, the components along the applied field for every moment are summed up and the volume of the system is determined using a lattice constant of 2.99Å.[29,36]

## III. RESULTS AND DISCUSSION

In this section, we present the results of our theoretical calculations for the magnetic properties of FeRh and compare them to both experimental data and other theoretical findings.

### A. Thick film Properties

Before we discuss the effects of thickness on the properties of FeRh, we provide a thorough analysis of the thick film properties of FeRh. To model a thick film material (where surface effects are negligible) it is sufficient to consider a system that is 12 nm thick, i.e. 40 ML of Fe, as the material properties no longer change as more layers are added.

### 1. M versus T in Thick Films

With the chosen parameters, a phase transition from AFM to FM upon heating is observed at around 370 K, with a thermal hysteresis of about 30-40 K, and Curie temperature of 680 K. The results, in good agreement with many experiments [24,26,33,45,46] and theories [13,17,31] are shown in Fig. 3, where the magnetization is plotted as a function of temperature for both heating and cooling. The calculation is done at a low field of H = 1 kOe, so as to break the symmetry of the magnetic system.

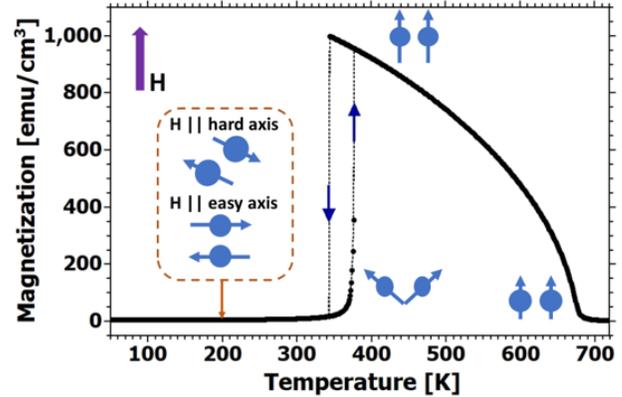

FIG. 3 Temperature dependence of the magnetization with an applied field of 1 kOe clearly shows the AFM (M=0) to FM phase transition and Curie temperature, as well as a thermal hysteresis of about 30- 40 K. The alignment of the two sublattice moments is shown by schematics in each phase of the system, with the applied field aligned along the easy axis or hard axis. The moments stay in the surface plane due to the demagnetization field, even with a relatively large out of plane anisotropy.

Schematic insets in Fig. 3 show the orientation of the spins on the two sublattices at different temperatures, which we explain now. At high temperatures the system is FM and the Fe moments are aligned with the external field. As the temperature is reduced, the thermal averaged size of the Fe moments increases. A further reduction in temperature causes a transition to an AFM state, where the moments' alignment depends on the orientation of the applied field with respect to the anisotropy axes. As can be seen in the insets of Fig. 3, at very low temperatures alignment of the applied field along the hard anisotropy axis produces moments that are antiparallel and lying at a small angle from an easy axis. If instead the applied field is along an easy axis, the moments lie perpendicular to H along the other easy axis. Upon heating from a low temperature, the system briefly enters a canted state before transitioning back to the ferromagnetic phase. With stronger applied fields, greater than 1 kOe, the strength and orientation of the applied field notably affects the alignments of the moments at low temperatures and consequently the AF-FM transition. This is discussed in detail later.

It should be noted that our $M_s$ is between 250-300 emu/cm³ lower than most measured values[16,24,35,45] in the FM state. As mentioned previously, the Rh moments are neglected in our model, therefore our magnetization is calculated considering the Fe moments only. At the transition, the additional magnetization from the neglected Rh atoms [31] (per unit cell there is one Rh atom with a moment of 0.9 $\mu_B$ in low temperature portion of the FM phase) causes the $M_s$ value here (a peak in Fig. 3 of 1000 emu/cm³) to be lower than that found in experiments (1250 emu/cm³).

## 2. M verses H in Thick Films:

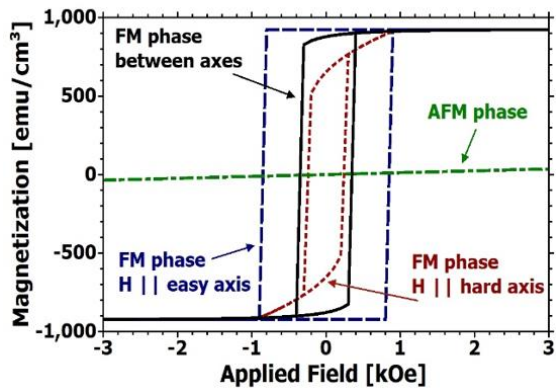

FIG. 4 Calculated in-plane magnetization at 340 K, with the magnetic field aligned along various axes for the FM phase. The results for the AFM phase, at 300 K, are also shown.

The calculated magnetization versus external field is plotted in Fig. 4. The field is applied along different orientational axes in the plane of the thick film. For FeRh in its FM phase, the results for field along an easy anisotropy axis (purple, long-dashed line), along a hard axis (red, fine-dashed line) and in-between the two (solid line) are shown. In addition, the result for the AFM phase (green, dot-dashed line) is shown, with negligible net moment. As previously mentioned, the orientation of the applied field with respect to the anisotropy axes has noticeable effects on the hysteresis in the FM phase as can be seen in Fig. 4. If the external field is aligned along the hard axis, the system displays a coercive field of about 250 Oe. As the angle of the applied field is moved away from the hard axis, the coercive field increases and reaches a maximum of about 1,000 Oe when aligned along the easy axis. The curvature in the hysteresis loop that is seen when the applied field is close to the hard axis is a result of significant canting of the moments as the field is increased.

In the low temperature region (AFM), the hysteresis curve (dot-dashed line in Fig. 4) shows no coercive field and very low magnetization. These results match well with experiments.[35,46] In the AFM state, the moments line up very close to an easy axis, but slightly canted toward the applied field, even for moderate applied magnetic fields. As a result, reversal of the magnetic field just moves the moments slightly and there is no energy barrier separating the +H configuration from the –H configuration, leading to the absence of a coercive field.

The M vs H curves for temperatures near the transition will be discussed below. We choose to align our external magnetic field between the easy and hard axes for all following calculations as this alignment matches most closely with experimental results.

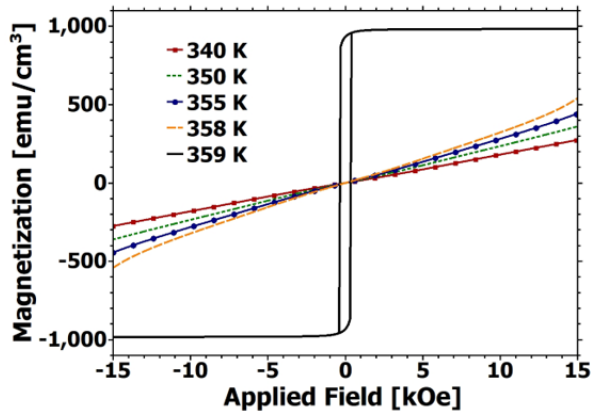

FIG 5. Calculated in-plane magnetization, M vs H curves at different temperatures near the phase transition.

Fig. 4 showed hysteresis curves for the AFM state at a particular temperature and for a limited field range. In Fig. 5 we expand the field range and examine the magnetization for different temperatures (340 K through 359 K) near the phase transition. The system is heated to the desired temperature with a 15 kOe field applied. Then the field is swept at a constant temperature. Noticeably, the magnetization increases with increasing temperature during the transition. This is because the spin-flop state becomes more canted before it transitions to the ferromagnetic state. The change in magnetization from 358 K to 359 K is abrupt because the AFM to FM transition has taken place. Note that this change is more gradual in real systems where AFM and FM domains can coexist across the transition.

Qualitatively our data matches that of experiments very well.[16,47] The discrepancy between theory and experiment is believed to be due to the lack of domains and defects in our simplified model. Stress and other interfacial effects due to a substrate,[17,24–26] in addition to defects in the material, are also believed to influence the transition, all of which are neglected in our model.

As mentioned earlier, we use a modified demagnetizing field of $3\pi M$ to include the large out-of-plane anisotropy as well as the true shape demagnetization factor. We justify this field by comparing theory results to experimental data. We compare the out-of-plane calculated magnetization curves with experimental data in Fig. 6, where different sized effective demagnetization factors (in CGS) are used, and the results compared against experimental out-of-plane data from Lu *et al.*[43] As can be seen in the figure the $3\pi M$ demagnetizing factor (green, dashed line) works reasonably well for fields below 6 kOe, to match the experimental results (blue dots). At higher fields we see a deviation which can be attributed to the discrepancies between our model and the true physical system as we neglect the Rh contribution, surface anisotropy, impurities, and domains. The in-plane (parallel) results are also presented as a reference (solid line).

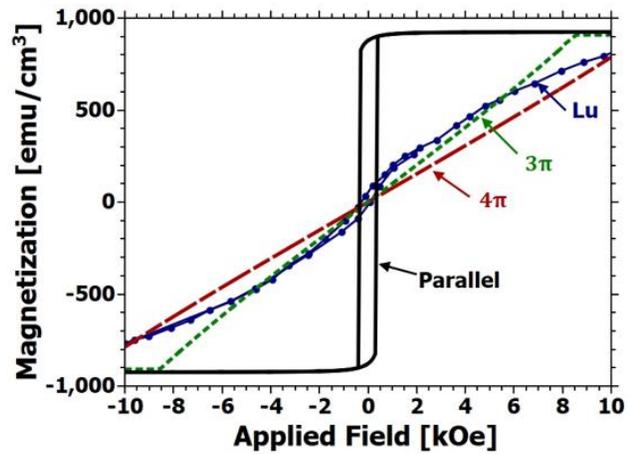

FIG 6. Calculated in-plane magnetization curves for different sized demagnetization factors at 410 K are shown (dashed lines). Experimental data from Lu (Ref. 42) are shown as dots for comparison. Calculated out-of-plane magnetization curve is given for reference (solid line)

### 3. Transition temperatures and thermal hysteresis as a function of applied field

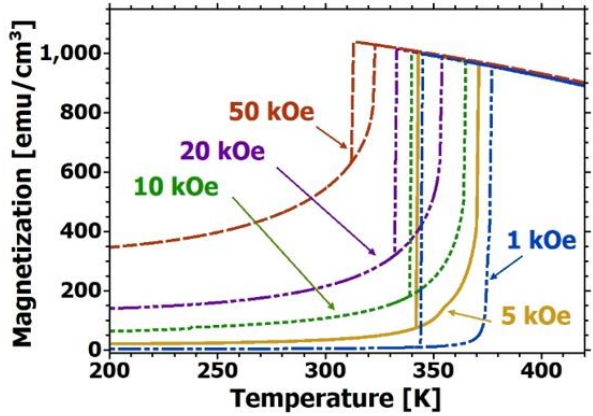

FIG 7. M vs T curves for various applied field strengths for a 40 Fe ML thick film. At high fields we see a field-induced magnetization in the AFM phase (low temperatures) plus a shift of the AFM to FM transition to lower temperatures.

The AFM to FM phase transition is expected to be sensitive to an applied magnetic field, shifting to lower temperatures because the field favors the FM state. We calculated the magnetization as a function of temperature for magnetic fields ranging from 1 kOe to 50 kOe, using a thick film system with 40 Fe ML. The results are shown in Fig. 7 for five different field values. As expected, the transition is shifted to lower temperatures in the presence of a strong magnetic field. This behavior is quantitatively the same in experiments[16,24,26,33,48], where the transition shifts to lower temperatures by about 20-30 K as the field is increased from 1 kOe to 20 kOe (see our purple line in Fig. 7).

Thermal hysteresis is found near the phase transition with a typical width of about 30-40 K for lower fields [Fig. 7]. This width decreases with increasing applied fields, a behavior which is also seen in Han *et al.*'s experiment[33] and which is typical for a system with thermal hysteresis,[49,50] as the large fields force the magnetization to orient in a particular direction.

An additional feature in Fig. 7 is the presence of a field-induced magnetization in the AFM phase. The field-induced magnetization can be understood by noting that the configuration is actually a spin flop state, where the moments on the Fe sublattices are mostly antiparallel but are canted toward the external field. This is shown in Fig. 8 with illustrations of the moments for the different anisotropy orientations, namely H applied along the (a) hard axis and (b) easy axis. As the applied field is increased, the canting becomes larger, leading to the increase in magnetization. This behavior is quantitatively consistent with that of that in experiments[26,33] with the measured magnetization of the AFM system, with values just below 400 emu/cm$^3$ for applied fields of 50 kOe. Interestingly, some experiments have not shown an increase in magnetization in the AFM state for large fields.[16,24,48]

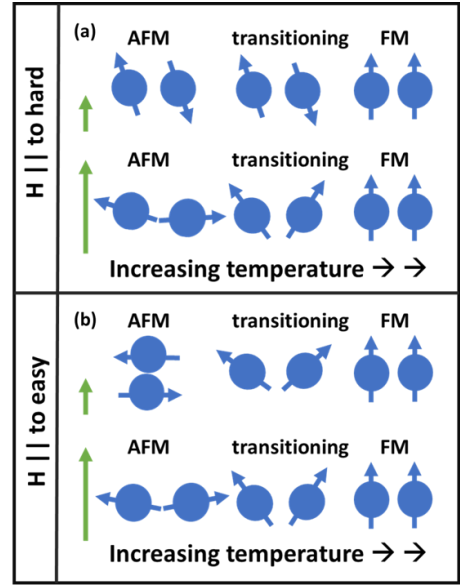

Alignment of the applied field with the anisotropy axes also plays a role in the orientation of the spins at low temperatures and at low fields as seen in Fig. 8. If the applied field is aligned along a hard axis (top panel (a)), different behavior is seen depending on the size of the external field. For low fields, less than 2,000 Oe, the spins are basically oriented along the easy axis. If the strength of the applied field is increased, the spins begin to rotate away from the easy axis and cant towards the applied field, but the spin flop state is not symmetric about the field direction because of the anisotropy. For larger fields, this rotation begins at lower temperatures and the canting becomes more significant, making the AFM-FM transition smoother. In contrast, if the applied field is aligned along an easy axis (bottom panel 8(b)), the magnetic moments lie along an easy axis perpendicular to the field at low fields. At higher fields a symmetric spin flop state can be seen with respect to the applied field.

FIG 8. Orientation of the Fe moments for different applied field strengths and directions as temperature is increased. The case of field applied along the hard direction is shown in the top panel (a) while along the easy direction is shown in the bottom panel (b).

## B. Thin film properties

Thin films are known to affect various properties of magnetic materials including hysteresis, magnetocaloric effect, magnetic state, and more.[17,33–35,50,51] However, only one paper[17] has calculated the effect of thickness on the FeRh system, when interface effects are important to consider. We model systems with thickness from a 2-3 ML to hundreds of ML to determine the effect of thickness on the phase transition, thermal hysteresis, as well as the magnetic structure of the system.

### 1. Transition temperatures and thermal hysteresis as a function of thickness

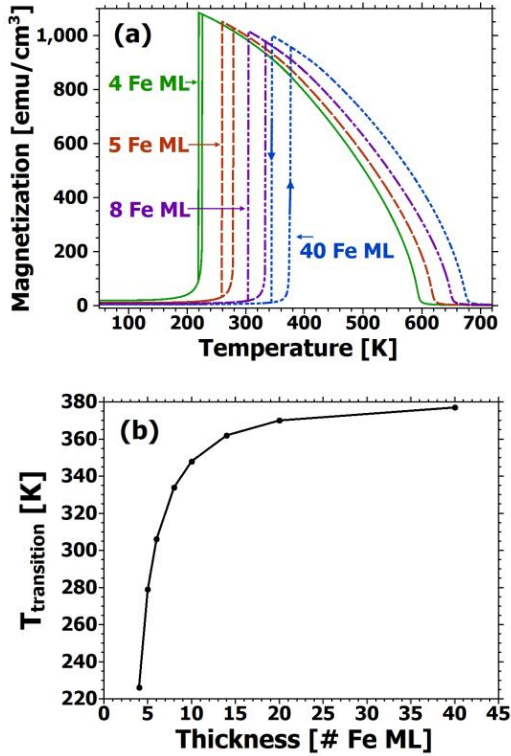

FIG. 9 (a) Magnetization as a function of temperature for thin FeRh films with different thicknesses. The applied field is 1 kOe. (b) The transition temperature as a function of thickness. We see a transition to bulk-like behavior near 20 ML.

Figure 9(a) shows the calculated thermal behavior of the magnetization as a function of temperature for FeRh films of different thicknesses, from 40 ML (fine-dashed, blue line) to 4 ML (solid, green line). As the thickness of the films decreases, the AFM to FM transition temperature decreases, as shown explicitly in Fig. 9(b). The transition temperature is defined as where the magnetization reaches its maximum value in the heating curve (see Fig 9(a)). Because the four-spin exchange interaction stabilizes the AFM state, this implies that the four-spin Hamiltonian is particularly sensitive to the presence of a surface. In an infinite, bulk, three-dimensional FeRh sample, the exchange interactions for each spin, both bilinear and four-spin, are completely satisfied in all directions. For thinner films, the interfaces have a significant contribution to the entire film. At the surface the four-spin contribution to the energy (which favors the AFM state) is decreased by one half, while the next nearest neighbor exchange is decreased by one third and the nearest neighbor exchange is only decreased by one sixth. Therefore, the thinner system remains ferromagnetic over a larger temperature range since the FM exchanges dominate over the four-spin exchange.

It is also evident from Fig. 9 that the thermal hysteresis decreases as the thickness decreases. Ostler[17] generated a numerical model which showed a decreasing thermal hysteresis width of about 15 K from an 8 nm to a 2 nm film. Our results show a decrease of about 20 K in width from 12 nm to 1 nm, which agrees well with this.

Surprisingly, experiments report contradicting results,[17,35] showing an increasing thermal hysteresis temperature width as thickness decreases. This discrepancy is significant yet can be explained by the lack of domains, defects, and a substrate in our simplified model, as discussed earlier. For example, the thinner the film, the more a substrate can influence exchange and therefore hysteresis effects.

Additionally, we find the AFM to FM transition is completely suppressed for films below 4 Fe ML (1.2 nm) in thickness, with the system remaining in a FM state across all temperatures. The lowest AFM to FM transition observed experimentally is at a thickness of 3 nm,[33] while the theoretical limit set by ab initio studies is 9 atomic layers when the film is Rh terminated.[22]

### 2. M vs H curves as a function of thickness

The thickness dependence of the M vs H curves at low temperature is explored in Fig. 10 for both the AFM (10a) and FM (10b) states. At low temperatures (AFM), with decreasing thickness, a larger magnetization is found. This emphasizes the importance of surface effects for thin films, allowing thinner films to have a larger degree of spin canting and therefore a larger moment at a given field and temperature.

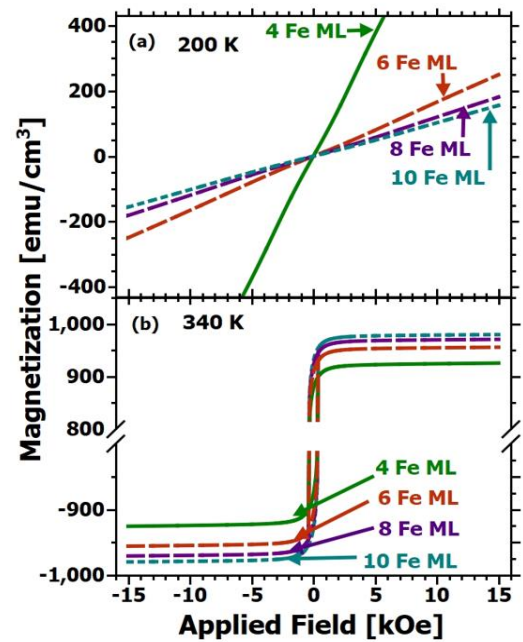

FIG. 10 M vs H curves for various thicknesses of thin FeRh films for both the (a) AFM and (b) FM states. Note the break in the magnetization scale in panel (b).

When the system is in the FM state, opposing behavior is seen (Fig. 10 b). Thinner films have a lower saturation magnetization than thicker ones after transitioning to the FM phase, at a constant temperature. At a given temperature in the FM phase, the moments in thinner films will have, on average, a smaller total exchange field, leading to a smaller thermal averaged magnetization. These results are in good qualitative agreement with experiments.[33,35]

It is also interesting to make note of the saturation of our system. While in the AFM state, our system does not saturate for the fields considered in Fig. 10(a), as also seen in Ref. 16 and Ref. 33. Rather it takes extremely large fields, on the order of 50-100 kOe, to saturate, which is in agreement with Ref. 30 and Ref. 35.

3. **Penetration depth of surface effects**

As noted earlier, the four-spin interaction is particularly sensitive to the presence of a surface, with the introduction of a surface favoring the ferromagnetic state. With this in mind, we study the typical penetration depth, defined below, for surface induced changes in this subsection. Figure 11 shows the canting angle, defined as the angle between the two Fe moments on different sublattices but the same layer, for various temperatures in the AFM phase.

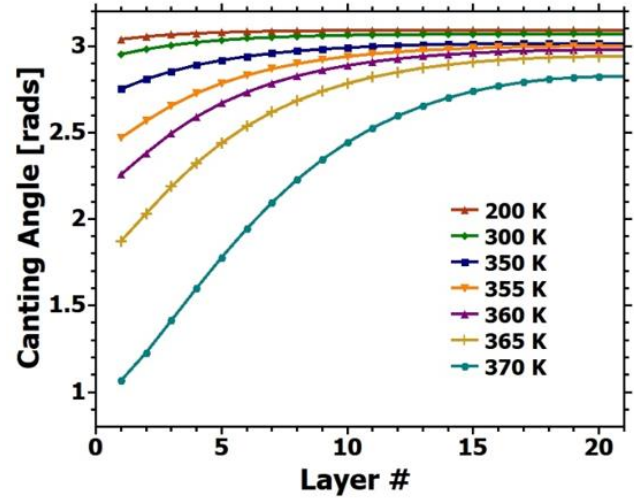

FIG. 11 Canting angle as a function of layer number for a 40 ML of Fe thick film in a 5 kOe applied field for various temperatures before and during the AFM to FM transition.

It is clear that as the temperature increases (bottom curves), the film becomes much softer, with canting increasing significantly near the surface and allowing the surface to have a deeper influence into the material's center. To quantify this, we define a penetration depth as the depth where the canting angle becomes 95% of the canting angle in the middle of the film. The penetration depth is then plotted as a function of temperature in Fig. 12 for 20 ML and 40 ML thick films. The penetration depth increases rapidly from 1 layer at low temperatures (below 310 K) to 35% of the way through the material right before the transition (370 K).

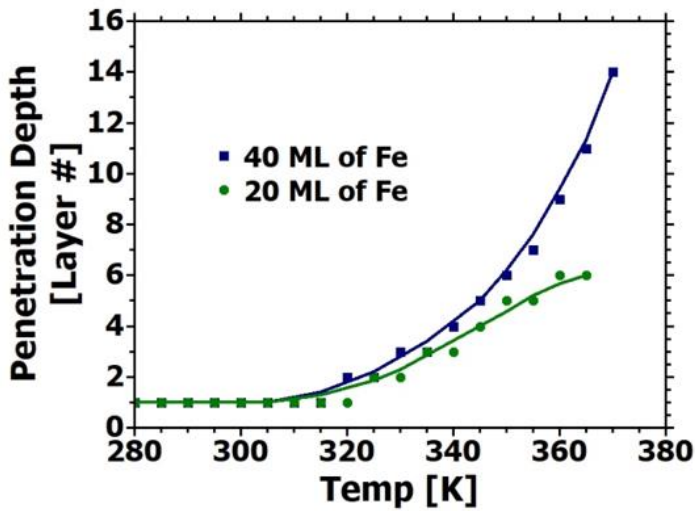

FIG. 12 Penetration depth as a function of temperature for a 40 ML and 20 ML film of FeRh in a 5 kOe applied field. Penetration depth is taken as the position where the canting angle is within 95% of the canting angle in the middle of the film.

The overall thickness of the film does not significantly affect the penetration depth for low temperatures, but deviations begin to arise at higher temperatures, which can be seen in Fig. 12, comparing the curves for 40 ML (square markers) and 20 ML (circles). For thicker films, the AFM state exists at higher temperatures. At these high temperatures, the system becomes softer, and the surface effects reach deeper into the film.

### IV. Conclusion

Using self-consistent local mean-field theory, a model is created to gain insight into FeRh, both in bulk and thin film form. With this model, a thorough theoretical description of the field and temperature phase diagrams is obtained, which are qualitatively in good agreement with a vast number of experiments and other theoretical results. In addition, our model allows for direct observation of each spin sublattice to visualize the system, specifically allowing one to see surface penetration depths and spin canting angles. This may lead to future predictions of interface effects when FeRh is combined with other magnetic materials, such as in a spin valve structure.[27] Our results validate that four-spin exchange interactions in competition with nearest-neighbor exchange and next-nearest-neighbor exchange are sufficient to reproduce the AFM-FM phase transition in FeRh. We note that a 3D Ising-like model (effectively huge in-plane anisotropy), initially used in Ref. 24 and Ref. 31, does not match experimental magnetization data.

We comment on the limitations and strengths of our model. First, our current model neglects the effect of the substrate on the magnetic properties of FeRh, which is known from experiment to play an important role.[17,24–26] Second, our model does not explicitly include the Rh and therefore does not differentiate between Fe terminated and Rh terminated films. While there have been some theoretical calculations on this issue[23], no experimental results have been reported. Third, as mentioned earlier, our model does not allow for lateral domains to form in FeRh. Atomistic calculations, in contrast, have taken this into account and the formation of domains smooths out transitions. Our model is, however, much less computationally demanding than atomistic simulations while reproducing qualitatively all thermal behaviors. Finally, we note that the calculations presented here use realistic anisotropy energy terms and strengths, unlike many previous theoretical works.[24,31]

A search for articles on FeRh shows there has been an exponential increase in the number of studies on this material in the past 20 years. Today, scientists and engineers are still interested in understanding the fundamental mechanism for the AFM to FM transition, plus in predicting the interface-induced and temperature-induced changes to the magnetic structure in order to incorporate FeRh into device applications at room temperature. The calculation presented here will aid researchers towards this goal.

# References

[1] M. Fallot, Ann. Phys. **10**, 291 (1938).

[2] M. Fallot and R. Hocart, Rev. Sci. **77**, 498 (1939).

[3] G. Shirane, R. Nathans, and C.W. Chen, Phys. Rev. **134**, A1547 (1964).

[4] Y. Liu, L.C. Phillips, R. Mattana, M. Bibes, A. Barthélémy, and B. Dkhil, Nat. Commun. **7**, 11614 (2016).

[5] M.P. Annaorazov, H.M. Güven, and K. Bärner, J. Alloys Compd. **397**, 26 (2005).

[6] M.P. Annaorazov, S.A. Nikitin, A.L. Tyurin, K.A. Asatryan, and A.K. Dovletov, J. Appl. Phys. **79**, 1689 (1996).

[7] L. Zsoldos, Phys. Status Solidi **20**, K25 (1967).

[8] M.R. Ibarra and P.A. Algarabel, Phys. Rev. B **50**, 4196 (1994).

[9] S.O. Mariager, F. Pressacco, G. Ingold, A. Caviezel, E. Möhr-Vorobeva, P. Beaud, S.L. Johnson, C.J. Milne, E. Mancini, S. Moyerman, E.E. Fullerton, R. Feidenhans'L, C.H. Back, and C. Quitmann, Phys. Rev. Lett. **108**, 087201 (2012).

[10] F. Bergevin and L. de Muldaver, J. Chem. Phys **252**, 1347 (1961).

[11] P.A. Algarabel, M.R. Ibarra, C. Marquina, A. Del Moral, J. Galibert, M. Iqbal, and S. Askenazy, Appl. Phys. Lett **66**, 3061 (1995).

[12] G. Ju, J. Hohlfeld, B. Bergman, R.J.M. van deVeerdonk, O.N. Mryasov, J.Y. Kim, X. Wu, D. Weller, and B. Koopmans, Phys. Rev. Lett. **93**, 197403 (2004).

[13] J. Barker and R.W. Chantrell, Phys. Rev. B **92**, 94402 (2015).

[14] J.-U. Thiele, M. Buess, and C.H. Back, Appl. Phys. Lett. **85**, 2857 (2004).

[15] J.U. Thiele, S. Maat, and E.E. Fullerton, Appl. Phys. Lett. **82**, 2859 (2003).

[16] J. Cao, N.T. Nam, S. Inoue, H. Yu, Y. Ko, N.N. Phuoc, and T. Suzuki, J. Appl. Phys **103**, 07F501 (2008).

[17] T.A. Ostler, C. Barton, T. Thomson, and G. Hrkac, Phys. Rev. B **95**, 64415 (2017).

[18] J.S. Kouvel, J. Appl. Phys. **37**, 1257 (1966).

[19] S.P. Bennett, H. Ambaye, H. Lee, P. Leclair, G.J. Mankey, and V. Lauter, Sci. Rep. **5**, 9142 (2015).

[20] P. Kushwaha, A. Lakhani, R. Rawat, and P. Chaddah, Phys. Rev. B **80**, 174413 (2009).

[21] J. Van Driel, R. Coehoorn, G.J. Strijkers, E. Brück, and F.R. De Boer, J. Appl. Phys. **85**, 1026 (1999).

[22] S. Lounis, M. Benakki, and C. Demangeat, Phys. Rev. B **67**, 094432 (2003).

[23] S. Jekal, S.H. Rhim, S.C. Hong, W.J. Son, and A.B. Shick, Phys. Rev. B **92**, 064410 (2015).

[24] S. Maat, J.U. Thiele, and E.E. Fullerton, Phys. Rev. B **72**, 214432 (2005).

[25] H. Kumar, D.R. Cornejo, S.L. Morelhao, S. Kycia, I.M. Montellano, N.R. Álvarez, G. Alejandro, and A. Butera, J. Appl. Phys. **124**, 85306 (2018).

[26] Y. Ohtani and I. Hatakeyama, J. Magn. Magn. Mater. **131**, 339 (1994).

[27] S. Yuasa, M. Nývlt, Katayama T., and Suzuki Y., J. Appl. Phys. **83**, 6813 (1998).

[28] R.Y. Gu and V.P. Antropov, Phys. Rev. B **72**, 012403 (2005).

[29] G. Shirane, C.W. Chen, P.A. Flinn, and R. Nathans, Phys. Rev. **131**, 183 (1963).

[30] J.B. McKinnon, D. Melville, and E.W. Lee, J. Phys. C Solid State Phys. **3**, S46 (1970).

[31] M.E. Gruner, E. Hoffmann, and P. Entel, Phys. Rev. B **67**, 064415 (2003).

[32] R.E. Camley, Phys. Rev. B **35**, 3608 (1987).

[33] G.C. Han, J.J. Qiu, Q.J. Yap, P. Luo, D.E. Laughlin, J.G. Zhu, T. Kanbe, and T. Shige, J. Appl. Phys. **113**, 17C107 (2013).

[34] J.M. Lommel, J. Appl. Phys. **37**, 1483 (1966).

[35] I. Suzuki, T. Koike, M. Itoh, T. Taniyama, and T. Sato, J. Appl. Phys. **105**, 07E501 (2009).

[36] V.L. Moruzzi and P.M. Marcus, Phys. Rev. B **46**, 2864 (1992).

[37] A. Szajek and J.A. Morkowski, Phys. B **193**, 81 (1994).

[38] O.N. Mryasov, Phase Transitions **78**, 197 (2005).

[39] J.-Y.P. Delannoy, M.J.P. Gingras, P.C.W. Holdsworth, and A.-M.S. Tremblay, Phys. Rev. B **72**, 115114 (2005).

[40] Y.O. Kvashnin, S. Khmelevskyi, J. Kudrnovský, A.N. Yaresko, L. Genovese, and P. Bruno, Phys. Rev. B **86**, 174429 (2012).

[41] N. Kazantseva, D. Hinzke, U. Nowak, R.W. Chantrell, U. Atxitia, and O. Chubykalo-Fesenko, Phys. Rev. B **77**, 184428 (2008).

[42] H.J. Schmidt and Y. Kuramoto, Phys. C **167**, 263 (1990).

[43] E. Mancini, F. Pressacco, M. Haertinger, E.E. Fullerton, T. Suzuki, G. Woltersdorf, and C.H. Back, J. Phys. D. Appl. Phys. **46**, 245302 (2013).

[44] K.L. Livesey and R.L. Stamps, Phys. Rev. B **81**, 064403 (2010).

[45] J.S. Kouvel and C.C. Hartelius, J. Appl. Phys. **33**, 1343 (1962).

[46] Y. Ding, D.A. Arena, J. Dvorak, M. Ali, C.J. Kinane, C.H. Marrows, B.J. Hickey, and L.H. Lewis, J. Appl. Phys. **103**, 07B515 (2008).

[47] W. Lu, Y. Wang, B. Yan, and T. Suzuki, J. Mater. Sci. **45**, 4919 (2010).

[48] S. Inoue, H.Y.Y. Ko, and T. Suzuki, IEEE Trans. Magn. **44**, 2875 (2008).

[49] R.E. Camley, W. Lohstroh, G.P. Felcher, N. Hosoito, and H. Hashizume, J. Magn. Magn. Mater. **286**, 65 (2005).

[50] A.L. Dantas, R.E. Camley, and A.S. Carriço, Phys. Rev. B - Condens. Matter Mater. Phys. **75**, 094436 (2007).

[51] F.C.M. Filho, V.D. Mello, A.L. Dantas, F.H.S. Sales, and A.S. Carriço, in *J. Appl. Phys.* (2011), p. 07A914.